\documentclass[]{kluwer}
\usepackage[dvips]{graphics}

\newcommand{\Ha}{H$\alpha$}
\newcommand{\Hd}{H$\delta$}
\newcommand {\Na} {NaD$_2$}

\begin{document}
\begin{article}
\begin{opening}
\title{Search for Microflaring Activity in the Magnetic Network }
\author{G. \surname{Cauzzi}}%\email{gcauzzi@arcetri.astro.it}}
\author{A. \surname{Falchi}}%\email{falchi@arcetri.astro.it}}
\institute{Osservatorio Astrofisico di Arcetri, L. Fermi 5, I-50125
Firenze, Italy}
\author{R. \surname{Falciani}}%\email{falciani@arcetri.astro.it}}
\institute{Dip. di Astronomia e Scienza dello Spazio, L. Fermi 5,
I-50125 Firenze, Italy}
\begin{ao}
A. Falchi\\
Osservatorio Astrofisico di Arcetri\\
L. Fermi 5\\
I-50125 Firenze\\
Italy 
\end{ao}

\begin{abstract}
We analyze the temporal behavior of Network Bright Points (NBPs)
searching for low-atmosphere signatures of flares occuring on the magnetic
network. We make use of
a set of data acquired during
coordinated observations between ground based
observatories (NSO/Sacramento Peak) and the MDI instrument onboard 
SOHO. 
Light curves in chromospheric spectral lines show only small amplitude temporal
variations, without any sudden intensity enhancement 
that could suggest the presence of a transient phenomenon such as a 
(micro) flare.
Only one NBP shows spikes of downward velocity, of the order of
2 -- 4 Km/s, considered as signals of compression associated to
a (micro) flare occurrence. 
For this same NBP, we also find a peculiar relationship between the
magnetic and velocity fields fluctuations, as measured by MDI. Only
for this point the B - V
fluctuations are well correlated, suggesting the presence of
magneto-acoustic waves propagating along the magnetic structure. This
correlation is lost during the compression episodes and resumes
afterward.
An A6 GOES soft X-ray burst is temporally associated 
with the downward velocity episodes, suggesting that this NBP is the
footpoint of the flaring loop. 
This event has  a total thermal energy
content of about $10^{28}$ erg, and hence belongs to the microflare
class.

\end{abstract}
\end{opening}

\section{Introduction}

High resolution observations taken by, e.g., YOHKOH and SOHO
during the solar minimum years have revealed small scale transient
phenomena occurring everywhere on the solar surface.
Their presence
could have important consequences on the issue of coronal heating:
depending on
their energy content and distribution, and rate of occurrence, they
could
provide enough energy to heat the corona (see reviews in \opencite{ulm91}).
Moreover, if these transient events represent the 
low energy tail of the broader solar flare family, their study could
more easily identify the ``basic'' flare mechanisms, not biased by
global collective effects as in the case of major solar flares.

Impulsive variations of the emission in the corona or transition region,
 that could be defined as low-energy flares,
have long been observed in both ``quiet Sun'' regions, within magnetic network
structures (\opencite{gol74}, \opencite{por87}), 
 and active regions \cite{shim95}. %(Shimizu, 1995). 
In this paper we will focus on the characteristics of 
the brightenings associated to the magnetic network.

\inlinecite{kru97} observed some small scale, transient 
soft X-ray brightening events associated to network magnetic bipoles.
These events are spatially and temporally associated with microwave emission,
and show several similarities with more energetic flares, such as the
same ratio between the total energy emitted in soft-X and radio
frequencies, etc.\ 
However, the total observed radiative losses for these events are more than one
order of magnitude smaller than previously reported X-ray events.
For these brightenings they introduced the term {\it soft X-ray network flares}.

Recent SOHO results, based on MDI, EIT, SUMER and CDS observations,
have further demonstrated the highly dynamic nature of the 
``quiet magnetic network''. 
Many observations confirm the existence of EUV brightening
events in small-size structures. 
\inlinecite{harr97} and \inlinecite{harr99} observed rapid variations in 
the emission of transition region
lines, associated with junctions of the network magnetic field. 
The emission characteristics of these brightenings, called blinkers,
suggest that they are a class of phenomena distinct from the network flares.
Spectrographic observations have shown bursts of highly Doppler-shifted
emission, called transition
region explosive
events (among others, \opencite{der94};
\opencite{inn97}; \opencite{cha98}).
\inlinecite{cha00} explored the relationships between blinkers and
explosive events. They concluded that although the two phenomena appear 
different because of the differences in size and duration, they seem
to be closely related to each other, for explosive
events occur at
the edges of blinkers and at the same time. 
Magnetic reconnection in different magnetic geometries is invoked
to explain the observed differences.

\inlinecite{ben99} and \inlinecite{kru00}
searched for EUV  and radio emission variations in 
non active regions associated to the magnetic network. They observed 
faint coronal and transition region fluctuations sharing some general
properties with more energetic flares (such as the delay between
emission originating at different atmospheric layers) and interpreted
them as  micro- or nano-flares occurring onto the magnetic network.

The investigations mentioned above refer predominantly to
coronal and transition region spectral signatures and to photospheric
magnetic field measurements. 
At chromospheric and photospheric levels the bright features 
associated to the magnetic
network do show intensity variations in many  spectral signatures 
(\opencite{lit93}; \opencite{cau00}).
However, the interpretation for these fluctuations is
still an open problem: they might be the result of magneto-acoustic waves
which propagate upward along the magnetic lines of force 
(\opencite{has99}) or they could represent an instability of the chromosphere
itself, as a response to micro or nano-flares processes (\opencite{han97}).
These fluctuations are usually studied in a {\it statistical} sense and
have never been put into direct relation with the network flares
identified at higher atmospheric layers.

In this paper we will discuss the search performed at chromospheric
and photospheric levels for flares occurring on the magnetic network.
Beside the customary analysis of the intensity and velocity variations, we also
investigate the changes of the magnetic flux density, and their relation with
the fluctuations of the velocity field.
We use a subset of the high spatial resolution observations 
obtained in August 1996 
during a coordinated observing program between ground based observatory
(``R.B. Dunn'' Solar Telescope at NSO/Sacramento Peak) and the Solar and 
Heliospheric Observatory (SOHO).

\section{- Observations and Data Reduction}

Table \ref{tbl-1} reports a summary of the observing setup, more accurately described
in \citeauthor{cau97} (\citeyear{cau97}, \citeyear{cau99}).
We give here only some short information
for the data used in this paper.
Monochromatic intensity images were obtained with the tunable
Universal Birefringent Filter (UBF) and the Zeiss Filter, at high
spatial and temporal resolution. Spectra have been acquired with the
Horizontal Spectrograph (HSG) around the CaII
K and H$\delta$ lines, setting the slit in different positions within the
field of view (FOV).
Onboard SOHO, the Michelson Doppler Imager (MDI, \opencite{scher95})
acquired data with an image scale of $0.605^{\prime\prime}/$pixel. Maps of
continuum intensity,
line-of-sight velocity and longitudinal magnetic flux were obtained in
the NiI 6768 \AA~line at a rate of one per minute for several hours.
The velocity images were available with a binned 2$\times$2 format.
The different spectral signatures allow a good coverage of the whole
photospheric - chromospheric region.

\begin{table}
\caption{Summary of the observational set-up. The last column gives the
temporal cadence of the corresponding spectral feature.}\label{tbl-1}
\begin{center}\scriptsize
\begin{tabular}{llllrr}
%\tableline
\hline
Instrument & FOV & Spat. resol. & Observing $\lambda$ (\AA) & FWHM (\AA)
& $\Delta$t (s) \\
\hline
%\tableline
% 
 & & & & & \\
UBF & 2$^\prime \times 2^\prime$ & $0.5^{\prime\prime} \times
0.5^{\prime\prime}$ & 5889.9 (NaD$_2$) & 0.2 & 12 \\
 & & & 5875.6 (HeI D$_3$) & 0.2 & \\
 & & & 6562.8 (H$\alpha$) & 0.25 & \\
 & & & 6561.3 (H$\alpha -1.5$ \AA) & 0.25 & \\
Zeiss & 2$^\prime \times 2^\prime$ & $0.5^{\prime\prime} \times
0.5^{\prime\prime}$ & 6564.3 (H$\alpha +1.5$ \AA) & 0.25 & 3 \\
White Light & 2$^\prime \times 2^\prime$ & $0.5^{\prime\prime} \times
0.5^{\prime\prime}$ & 5500 & 100 & 3 \\
HSG & $0.75^{\prime\prime}\times 2^\prime$ & $0.75^{\prime\prime}
\times 0.36^{\prime\prime}$ & 3904$-$3941 (CaII K) & 0.035 &  -- \\
 &  & $0.75^{\prime\prime}\times 0.13^{\prime\prime}$ & 4094$-$4108 
(H$\delta$) & 0.011 & --  \\
MDI &10$^\prime \times 6^\prime$ &$0.6^{\prime\prime}\times
0.6^{\prime\prime}$&  6768 NiI & 0.1 &60 \\
\hline
\end{tabular}

\end{center}
\end{table}

The data analyzed in the present paper were obtained on August 15th, 1996,
when we had a period of constant good seeing conditions
(better than 1$^{\prime\prime}$) all along one hour (15:15-16:05 UT).
The MDI data were considered for a longer interval, 14-17 UT.
A small Active Region (AR NOAA 7984), with a very low activity level, 
was present in the field of view. Some pores and a small spot were
the prominent structures in the white light and continuum images, 
and have been used for co-alignement of the whole
dataset. 
At each given time the alignment among the images acquired with
different instruments was better than about  $1^{\prime\prime}$.
We refer to \inlinecite{cau00} 
for details on FOV overlay 
and data reduction procedures.

\begin{figure}[t]  
\centerline{\includegraphics{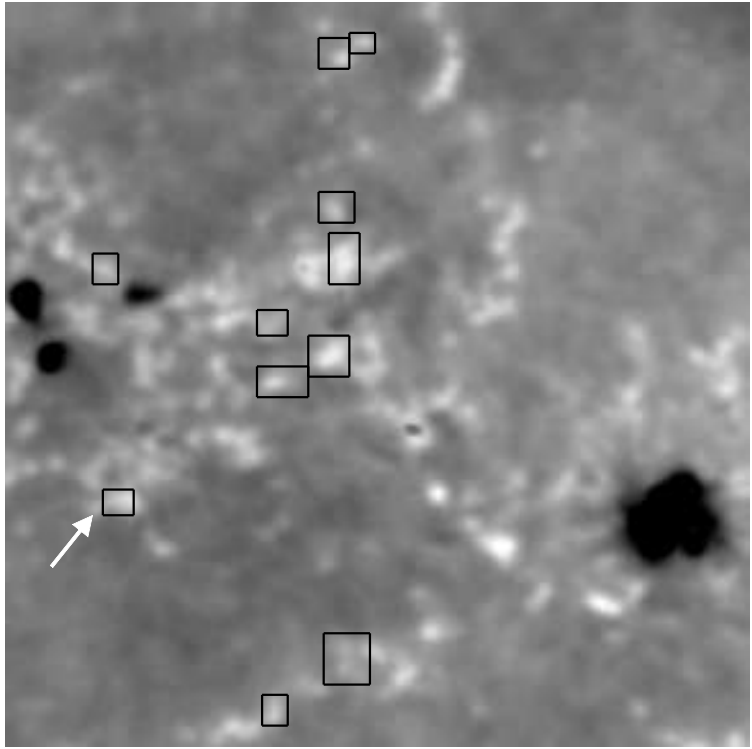}}
  \caption{ \Na~ image of the FOV, averaged over the period 15:15-16:05 UT,
  Aug. 15 1996. The bright pattern closely overlays the network magnetic
  structure (Cauzzi {\it et al.}, 2000). The black squares indicate the 11
  NBPs analyzed. The arrow indicates the peculiar NBP considered in the
  following.} \label{fig_fov}
\end{figure}

The magnetic network structure was clearly visible within the FOV in
several sets of images (MDI magnetic maps; \Na~ intensity; \Ha \ wings).
We searched for activity manifestations on several Network Bright Points
(NBP), sharing the following properties:  - bright
in the Ca II wings and in the Ca II 
K$_2$ peaks; - visible in the \Na~ images for about 1 hr; - spatially
coincident with magnetic structures.
We identified a total of 11 NBPs with the required characteristics,
displayed in Figure~\ref{fig_fov}.
We did not consider other bright points visible in the FOV,
since no corresponding CaII K spectra were available.

\section{Flaring Activity on the Magnetic Network}

\subsection{Intensity and velocity analysis}

As a first step, we searched for the presence of flaring episodes 
within the network  by analysing the intensity light curves,
obtained separately for each NBPs and each spectral signature.
Light curves were obtained setting an intensity threshold for each
signature, and averaging over the spatial pixels that
exceeded this threshold (more details can be found in \opencite{cau00}).
For the white light, NiI continuum and HeI- D$_3$ images 
no intensity threshold could clearly discriminate between NBPs and
surrounding areas,
within the limits of our statistical photometric
precision ($\approx$ 0.5\%).

\begin{figure}  
\centerline{\includegraphics{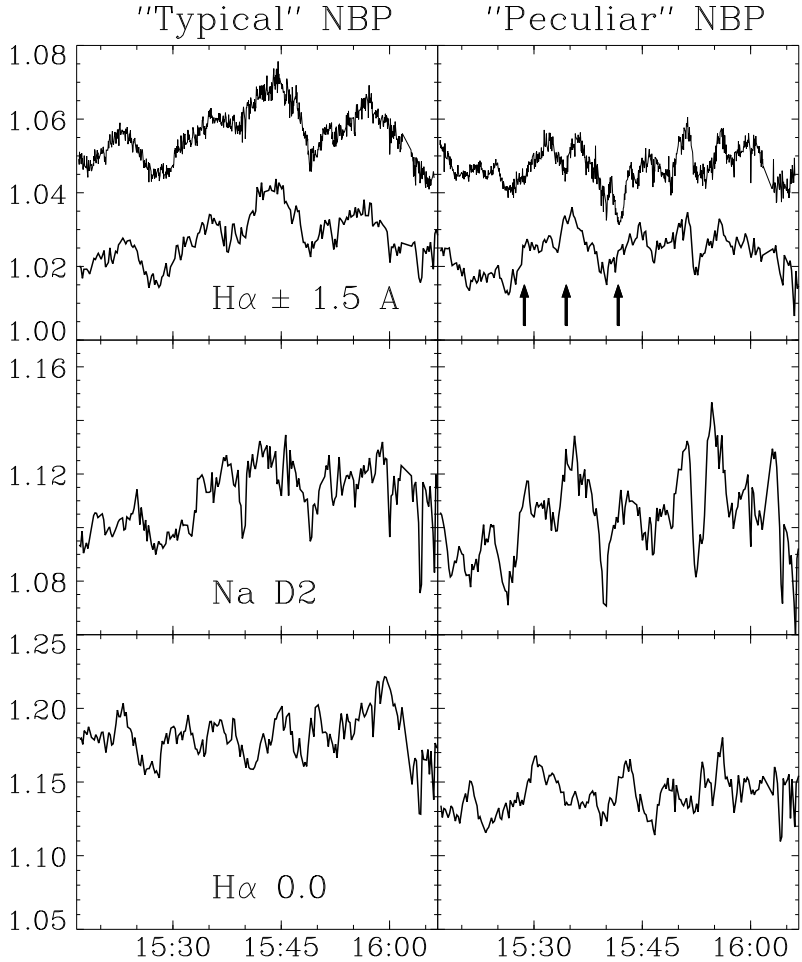}}
  \caption{Intensity  light curves, obtained in several spectral signatures, for
  two NBPs. The curves are normalized to the quiet average value and
  long period trends have been subtracted. In the first row the higher
  curve (shifted upward by 0.03 for clarity) represents the \Ha $+ 1.5$ \AA \ 
  emission. Left column: ``typical'' NBP with simultaneous intensity
  variations in the H$\alpha$~wings; right column: the only NBP
  showing non-correlated variations in the H$\alpha$~wings (indicated by
  arrows).}
  \label{light_curves}
\end{figure}

In Figure \ref{light_curves}, as an example, 
we show light curves of different signatures for two network bright points.
The variations of the emission are small in amplitude and
semi-regular in their temporal distribution. None of the analysed NBPs displays
any sudden intensity enhancement that could suggest the presence of a transient
phenomenon such as a micro-flare.
A search for the presence of regular
fluctuations has been performed in \inlinecite{cau00} and
reveals different temporal frequencies for different spectral signatures.

The fluctuations in different spectral signatures are 
not clearly in phase with each other (see Figure \ref{light_curves}).
An exception is represented by the H$\alpha \pm1.5$~\AA~wings light
curves (Figure \ref{light_curves}, left).
The evolution of most NBPs in the red and blue wings
of H$\alpha$ is very similar, with simultaneous intensity variations 
of the same amplitude, between 0.5 and 1.5\% of their average values. 
This similarity indicates that velocity fluctuations do not play a role in
the observed intensity variations, and  that these are most likely related to 
temperature or density fluctuations.

However, one of the analysed NBPs displays out-of-phase 
intensity variations  in the wings of H$\alpha$, that could be 
explained by an asymmetric absorption profile (see Figure \ref{light_curves}, 
right). 
The hypothesis of an \Ha~ absorption profile seems reasonable,
since the light curves obtained in the \Ha~ center do not show any
particular enhancement at any time and the \Hd~ spectra acquired on
several NBPs at different times always show an absorption profile.

Interpreting the
difference in intensity as due to a velocity field (as commonly
done in dopplergrams), we derived the line-of-sight velocity
corresponding to the layers that contribute to the H$\alpha$ wings. 

\begin{figure}[h]
\centerline{\includegraphics{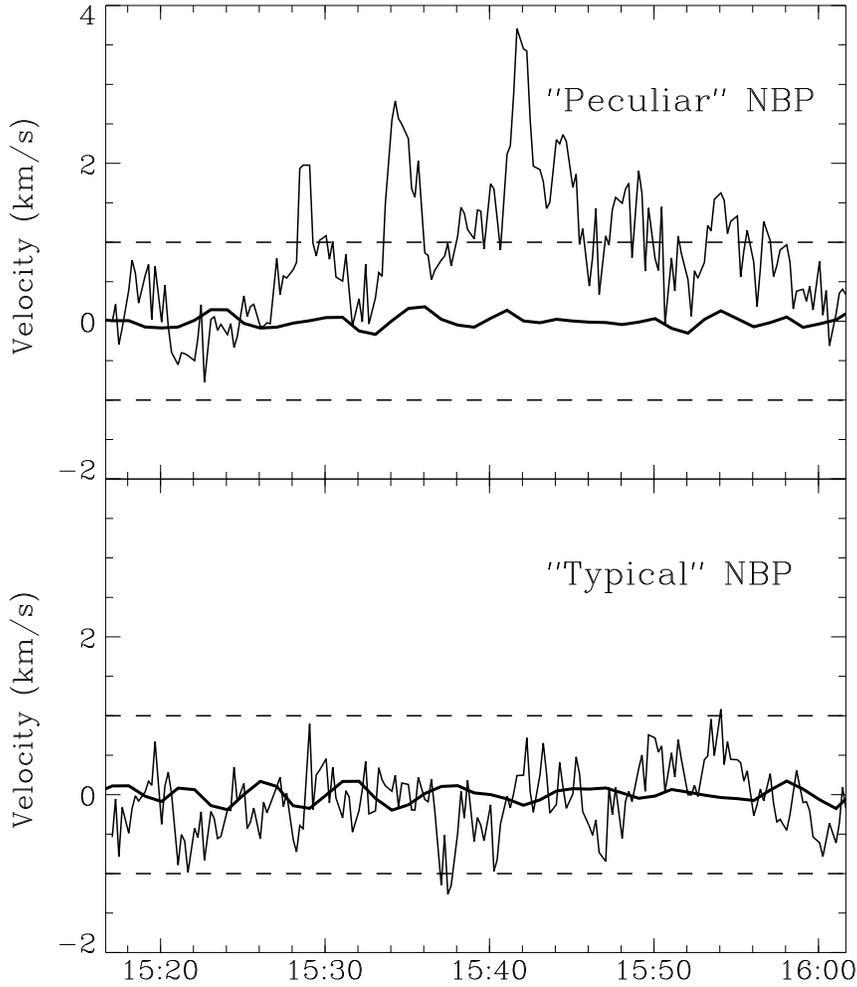}}
  \caption{ (Top) Line-of-sight velocity distribution (positive
  values indicate downward motions) for the particular NBP
described in text.
(Bottom)
Line-of-sight velocity distribution for one typical NBP. 
%showing that the signal is zero within the uncertainty.
Thin line: \Ha~wings velocity, with the error bar indicated by the dashed line. 
Thick line: MDI-NiI velocity, $\pm0.2$ km/s.
} 
\label{ha_vel}
\end{figure}

In Figure \ref{ha_vel} we
show this velocity as a function of time for this particular
NBP (top panel) and for one ``typical'' NBP (bottom panel). 
The peculiar NBP displays three spikes of
downward velocity 
within the interval 15:25 -- 15:45 UT, with a typical lifetime 
of about 1.5 minutes.  The amplitude of these spikes is well above the
error 
estimated at about 1 km/s.

A downward motion measured in H$\alpha$ or other chromospheric
lines represents a typical 
signature during the impulsive phase of a flare
(\opencite{ich84}).
This motion is due to the sudden compression of the
lower atmosphere caused by either a particle beam or a conduction front. This
has been proved also for small size events, such as the microflares described
by \inlinecite{can87} 
as the H$\alpha$ counterparts of hard X-rays burst.
Hence, we can suppose that these downward velocity episodes
represent the signature of the occurrence of flare processes within
this peculiar NBP.

The MDI velocity fluctuations measured in this area (Figure \ref{ha_vel}, 
thicker line)
are comparable with those registered in any other NBP and, 
in particular, do not show any peak
in correspondence with the downward motions registered in the \Ha~wings. 
Probably, within the flare driven chromospheric
condensation, the \Ha~wings originate from 
higher atmospheric levels than the NiI line
used by MDI, and are more sensitive to the disturbances introduced by a flare.

\subsection{Velocity and magnetic field correlation}

\begin{figure}[] 

\centerline{\includegraphics{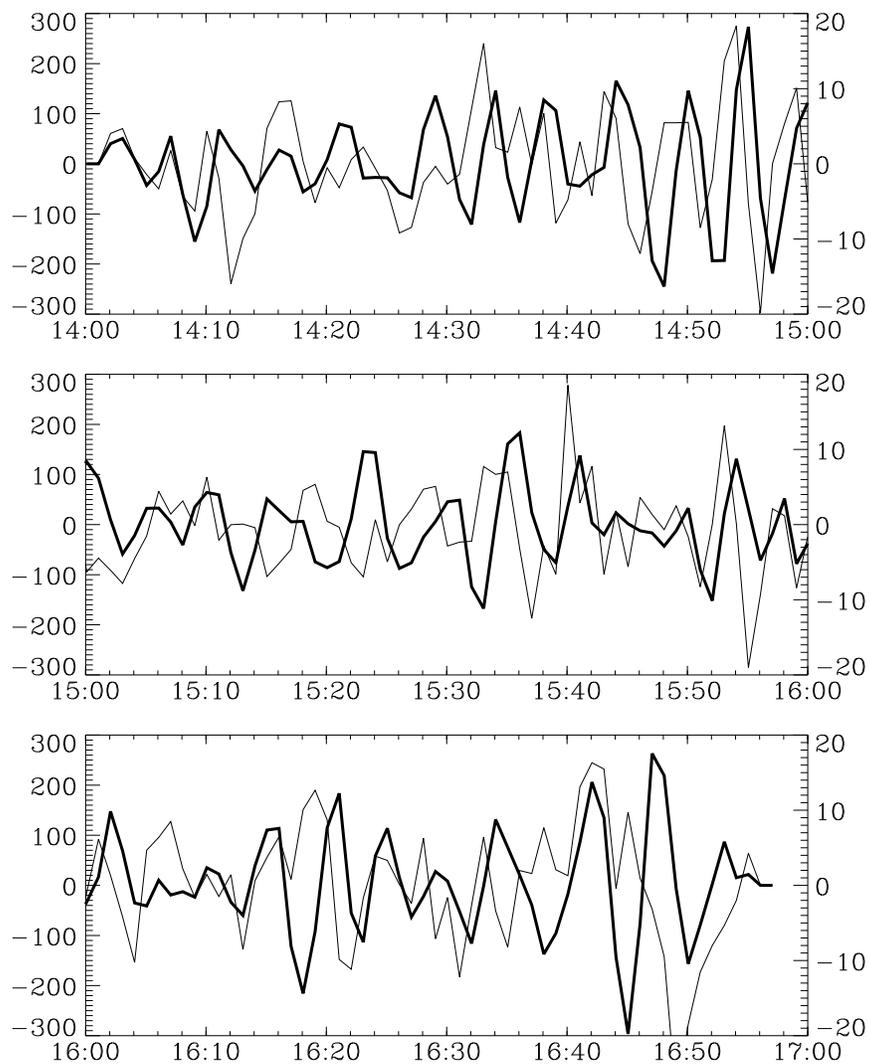}}
\caption{Comparison between the fluctuations of the MDI line-of-sight velocity 
(thick line, scale on the left in m/s) and of the longitudinal magnetic flux density 
(thin line, scale on the right in G). For both quantities the average value and 
the long period trends have been subtracted.}
\label{vel_B}
\end{figure}
This NBP coincides on the MDI maps with a magnetic structure of negative
polarity with a flux density smoothly increasing in time from 150 to 250 G.
Some very weak structures of opposite polarity, barely above the
noise (15 G), are present in the surroundings.

In general, within the NBP areas both the magnetic flux density (B) 
and the line of sight
velocity (V) fluctuate with a rough periodicity of 5 minutes, 
but these fluctuations do not
appear to be related to each other (see \opencite{cau00} for further
details).
For this particular NBP, the B and V fluctuations are instead well 
correlated for long
periods of time, 14:00 - 15:10 UT  and 15:40 - 17:00 UT (see Figure
\ref{vel_B}).
However, during the interval 15:10 - 15:40, when also the \Ha~downward 
motions are measured (see Figure \ref{vel_B}), their correlation gets
smaller.

\begin{figure}[] 
\centerline{\includegraphics{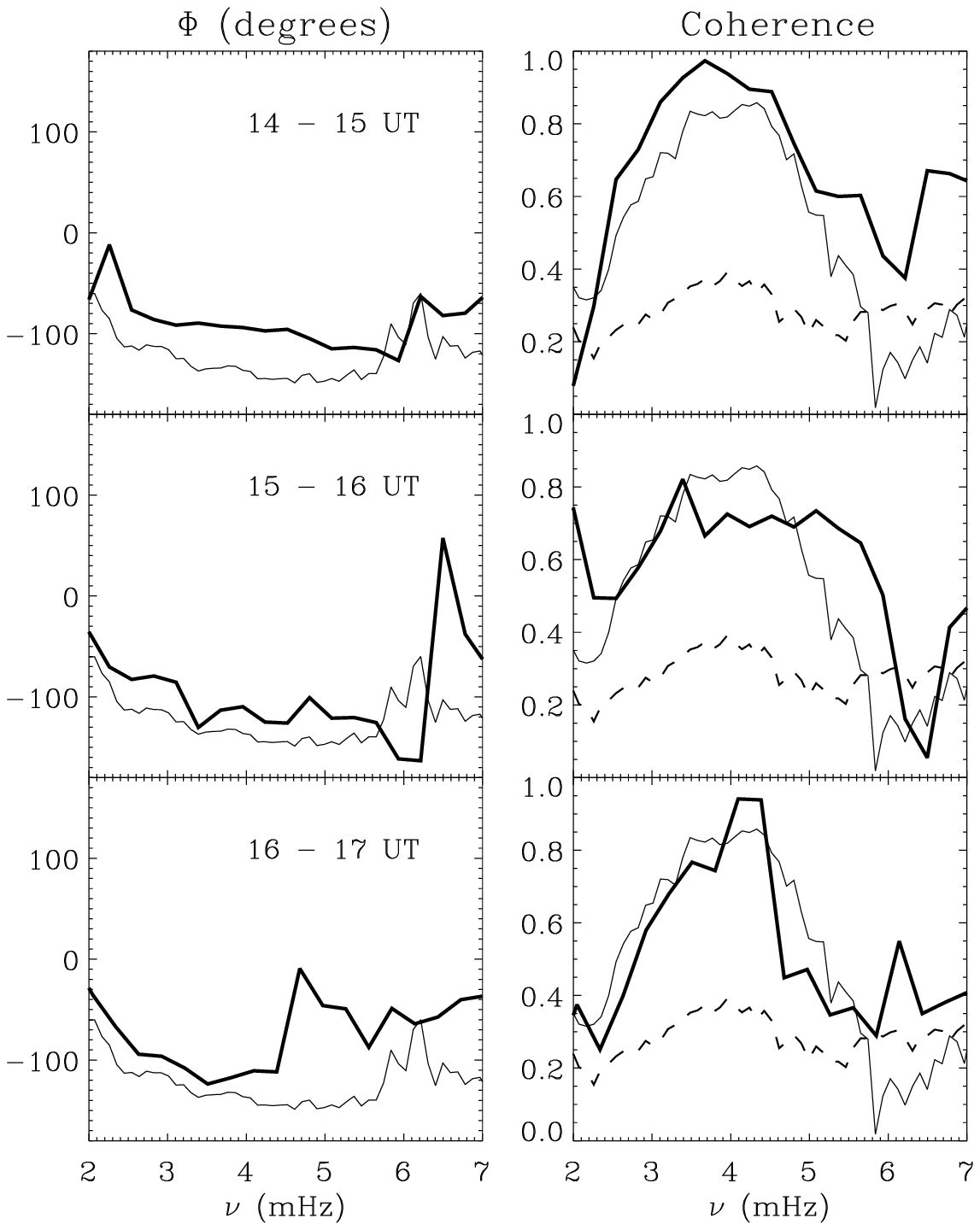}}
\caption{B - V phase difference (left) and coherence (right) spectra. 
%for the {\it flaring} NBP. 
From top to bottom the spectra (thick line) are shown for three consecutive time
windows: 14 - 15, 15 - 16 and 16 - 17 UT. The thin line represents the
average over the whole 3 hour period for the {\it flaring} NBP, while the
dashed line is an average for the remaining 10 NBPs.}\label{phas_cohe}
\end{figure}

These characteristics are confirmed by a more accurate analysis, that makes
use of the phase
difference ($\Phi$) and the coherence ($ C$) spectra in the Fourier domain 
for the pair B - V  (see \opencite{edmwe72} 
for definition of $\Phi$ and $C$).
Basically, the phase difference provides a measure of the delays between periodic signals,
while the coherence gives the persistence of this phase difference. 
Reliable values of phase difference can be given only if C has values higher than about 0.9.

In Figure \ref{phas_cohe} (left) we show the phase difference B - V for the peculiar NBP,
as a function of the
temporal frequency, computed for the whole observing period and for three
separated periods of one hour each. For this calculation we considered
positive the upward velocity, as it is customary.  The corresponding coherence 
spectra are shown in Figure \ref{phas_cohe}  (right). 
The coherence spectrum shows a peak of about 0.95 at the frequency
corresponding to 5 minutes oscillations in the time
intervals 14 - 15 UT and 16 - 17 UT. This maximum is lowered
to about 0.7-0.8 during
the period 15 - 16 UT, when B and V lose their correlation (Figure
\ref{vel_B}). 
At the frequency of the maximum coherence the
phase difference is of about $-90^{\circ}$, with the upward velocity V leading B. 
This is consistent with the presence of
propagating magnetoacoustic waves within this magnetic area 
(\opencite{ulr96}), 
possibily excited by granular buffeting of the magnetic elements
(\opencite{has99}).
For all the other NBPs the B - V coherence spectra assume
very low values (Figure \ref{phas_cohe}, dashed), 
and hence no phase difference analysis can be performed.

The lack of coherence between B and V fluctuations in the majority of
NBPs could be explained by the magnetic field configuration in the
network. As shown for example 
by \inlinecite{schri97}, 
the network magnetic cells
consist of a large number of very small concentrations of mixed
polarities, connected mostly via short loops. In such small structures,
wave reflections or distortion can destroy the coherence between B and V
fluctuations, not
allowing the persistence of any phase difference. The particular
NBP where we observe a definite phase difference, might instead be connected to 
the opposite polarity via a longer loop that hence allows wave
propagation. 

As said above, in this NBP the coherence between B and V is 
reduced during the interval 15:10 - 15:40 UT (Figure \ref{vel_B}).
The temporal coincidence with the \Ha~downward motions suggests that the
same phenomenon can be responsible also for this disturbance.
If the downward motions measured in \Ha~ are a signature of a flare, one
can suppose the occurrence of magnetic reconnection processes involving this
longer loop and having strong effects on the footpoint(s). 
During a reconnection process,
the magnetic topology changes abruptly and any
oscillations regime of B would be strongly perturbed. 
As a result, the wave propagation will be disrupted and the B - V 
coherence reduced. The phase difference between the fluctuations could then 
be re-established once the perturbation is over.
The MDI maps show some weak magnetic structures of opposite polarity
in the vicinity of this NBP within the considered temporal
interval. Any of these structures could be involved in the reconnection
process, but the complexity of the ``salt and pepper'' magnetic 
pattern makes it impossible to identify the particular feature involved.

\subsection{Soft X-ray burst}

The scenario outlined in the previous sections of a magnetic
reconnection and a sudden compression at low atmospheric layers
suggests the possibility of associated emission coming from higher
atmospheric layers.
Hence, we searched for coronal signatures temporally and/or spatially
associated with these signals from the lower atmosphere. 
Unfortunately, during the
considered time lag YOHKOH was in its night-time and 
only SXT full-disk frames were available before and after this interval.
Two  active regions were present on the disk: NOAA 7982 at the west
limb and NOAA 7984, visible in our FOV. An accurate analysis of
the images available for both regions, for two hours before and one
hour after, reveals that the SXR emission of NOAA 7982 was decreasing very
slowly during the whole period, while NOAA 7984 showed global intensity
fluctuations of about $\pm$ 2\%/hour.
This suggests that only the region
present in our FOV can show some episodes of activity.

\begin{figure}[h] 
\centerline{\includegraphics{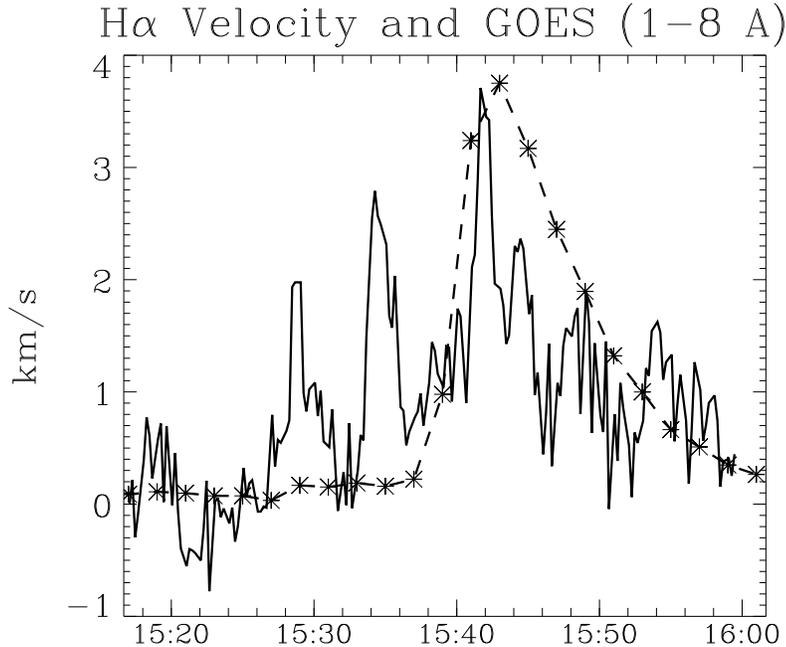}}
\caption{SXR emission from GOES 1--8 \AA~band (relative units), 
superimposed to \Ha~ line-of-sight velocity.}\label{soft_vel}
\end{figure}

GOES recorded an A6 burst in both
channels (0.5-4 and 1-8 \AA), with maximum peak at 15:43 UT. The temporal
evolution of the SXR emission in the 1-8 \AA~ band is shown in 
Figure \ref{soft_vel}, superimposed to the line-of-sight velocity outbursts 
deduced from the \Ha~wings. 
The temporal coincidence between the two curves is good.  The first two
velocity spikes, of small amplitude, correspond to a modest increase
of the SXR emission (build-up phase), while the third, larger one, 
corresponds to the peak of the GOES curve. 
If we accept the suggestion
that only the AR present in our FOV can display activity, 
this particular NBP most probably is the footpoint of a flaring
loop.

Following \inlinecite{tho85},
from the ratio of the GOES channels
power fluxes (0.5$-$4 \AA / 1$-$8 \AA), we can estimate an electron temperature 
T$_e$ of about   4.5~ MK and an emission measure  of about $ 10^{47}$ cm$^{-3}$. 
Assuming a mean electron density of $10^{10}$ cm$^{-3}$, 
(typical of dense coronal loops) we can
estimate a mean volume of $10^{27}$ cm$^3$ for the plasma emitting 
the measured SXR burst.
A cylindrical coronal loop, with a constant diameter of
5$^{\prime\prime}$ (somewhat larger than this NBP to take into
account the possible size increase of magnetic structures at coronal levels), 
would have a total length of $8\times10^4$ Km, with
a total thermal energy content of about $10^{28}$ erg.
Hence, in a completely independent way, we again suggest the presence of a
long magnetic loop with a footpoint in a network bright area.
The values found for T$_e$, for the EM and for the associated thermal
energy for this SXR event are consistent with those quoted by 
\inlinecite{fel96}
 for microflares of A6 GOES type. We can then conclude that
this event can be considered as a real {\it network microflare}.

\section{Conclusions}

We searched for the presence of chromospheric and photospheric
signatures of low-energy flares occurring on the magnetic network.
Eleven NBPs, each one corresponding to a patch of definite
polarity, were selected in a 2$^\prime\times 2^\prime$ field of view, where 
also a small and low-activity region was present. The intensity fluctuations 
of several spectral signatures observed for the NBPs
do not show any impulsive episodes
typical of the flare occurrence. The general characteristics
found for these NBPs do not differ from the ones derived in 
quiet regions (\opencite{cau00}).
Only for one NBP we found two peculiar characteristics: 
\begin{itemize}
\item[--] The intensity
fluctuations in the \Ha~wings ($\pm$ 1.5 \AA~) are not in phase for
about 20 minutes. From the difference of the intensity of the \Ha~wings
we derived the corresponding velocity and found that
three spikes of downward motions
(respectively of 2.0, 2.7 and 3.7 $\pm$ 1.0 Km/s) occurred within this
interval.
Plasma motions of this kind are generally  considered as a signature of sudden
compression of the lower atmosphere, typical of flares.

\item[--] In the Fourier domain, at the frequency corresponding to 
5 minutes oscillations, the MDI magnetic
and velocity fluctuations in this NBP show
a coherence spectrum with a high value ($\approx 0.8$, see Figure
\ref{phas_cohe}) when compared to the
average for all the other NBPs ($\approx 0.4$).
Analyzing the coherence spectrum for 3 separate periods,
we found that before and after the downward velocity episodes, 
the coherence has an even higher value of about 0.95, while just during the
sudden compression episodes, the coherence is lowered to 0.7.
We suggest that a reconnection process, that might be responsible for 
the sudden compression
indicated by the three spikes of downward velocity, might also destroy
the coherence of
magneto-acoustic waves propagating along the magnetic field lines of
a long loop connecting this NBP with a far opposite polarity.
\end{itemize}

GOES recorded an A6 burst in both channels, that temporally coincides
with the stronger downward velocity spike. We
derived  $T_e = 4.5$~ MK and EM $ \approx 10^{47}$ cm$^{-3}$ for a
coronal loop with a constant diameter of
5$^{\prime\prime}$ and  a total length of about $8\times10^4$ Km. 

The general picture emerging from the data is hence the following: the
particular network bright point in exam is connected via a long magnetic
loop to a far opposite polarity structure. In ``quiet'' conditions,
magneto-acoustic waves can propagate along this loop as derived from the
$-90^{\circ}$ phase difference measured between B and V fluctuations. 
Magnetic reconnection involving this loop disrupts the coherence
between these fluctuations, and produces the compression episodes
observed in the lower atmosphere. At coronal levels the simultaneous
increase in SXR emission might identify the corresponding chromospheric 
evaporation. 
All this evidence points toward the occurrence of a real ``network
flare''. Contrary to more energetic flares, for such an event 
a sudden intensity increase does not represent a characteristic signature
in the low atmosphere. Instead, downward motions seem to be a common
property for both larger and smaller flares. An intriguing
characteristic of this particular flare is given by the disruption of
the coherence between the B and V fluctuations. This is the first time
that such a property is observed, so it would be premature to comment on its
relevance as a flare signature. Further investigation on the subject is
desirable.

\begin{acknowledgements}
 The authors express their warmest thanks to the
MDI team (P.I. P.H. Scherrer) 
for the efficient support during the observing run.
The support of the NSO/Sacramento Peak "R.B. Dunn" Tower staff was
essential for getting good data.
The authors would like to thank Drs. E. Landi and M. Velli for
helpful collaboration and stimulating discussion. 
\end{acknowledgements}

\end{article}

\begin{thebibliography}{99}

\bibitem[\protect\citeauthoryear{Benz and Krucker}{1999}]{ben99} 
Benz, A.O. and Krucker, A.: 1999, {\it Astron. Astrphys.} 
{\bf 341}, 286
\bibitem[\protect\citeauthoryear{Canfield and Metcalf}{1987}]{can87}
Canfield, R.C. and Metcalf, T.R.: 1987, 
{\it Astrophys. J.} {\bf 321}, 586 
\bibitem[\protect\citeauthoryear{Cauzzi {\it et al.}}{1997}]{cau97} 
Cauzzi, G., Vial, J.C., Falciani, R. and Falchi, A.: 1997, in
B. Schmieder, J.C. del Toro Iniesta, M. V\'azquez (eds),
ASP Conf. Ser. Vol. 118, p. 309, 
\bibitem[\protect\citeauthoryear{Cauzzi {\it et al.}}{1999}]{cau99}
Cauzzi, G., Falchi, A., Falciani, R. and Vial, J.C.: 1999, in
A. Wilson (ed.), {\it Magnetic Fields and Solar Processes}, 
9-th EPS Solar Meeting, Florence, ESA SP-448, p. 685
\bibitem[\protect\citeauthoryear{Cauzzi, Falchi, and Falciani}{2000}]{cau00} 
Cauzzi, G., Falchi, A. and Falciani, R.: 2000, {\it Astron.
Astrophys.} {\bf 357}, 1093
\bibitem[\protect\citeauthoryear{Chae {\it et al.}}{1998}]{cha98} 
Chae, J., Lee, C.-Y., Wang, H., Goode, P. R., 
Sch\"{u}hle, U.: 1998, {\it Astrophys. J.} {\bf 497}, L109
\bibitem[\protect\citeauthoryear{Chae {\it et al.}}{2000}]{cha00} 
Chae, J., Wang, H., Goode, P.R., Fludra, A., Sch\"{u}hle,
U.: 2000, {\it Astrophys. J.} {\bf 528}, L119
\bibitem[\protect\citeauthoryear{Dere}{1994}]{der94} Dere, K.P.: 1994, {\it Adv. Sp. Res.} {\bf 14}, no. 4, 13
\bibitem [\protect\citeauthoryear{Edmonds and Webb}{1972}]{edmwe72} 
Edmonds F.N., jr. and Webb C.J.: 1972, {\it Solar Phys.} {\bf 22}, 276
\bibitem[\protect\citeauthoryear{Feldman, Doschek, and Behring}{1996}]{fel96}
Feldman, U., Doschek, G.A. and Behring, W.E.: 1996, 
{\it Astrophys. J.} {\bf 461}, 465
\bibitem[\protect\citeauthoryear{Golub {\it et al.}}{1974}]{gol74} Golub L., Krieger, A.S., Silk, J.K.,, Timothy, A.F.,
Vaiana, G.S., 1974, {\it Astrophys. J.} {\bf 189}, L93
\bibitem[\protect\citeauthoryear{Hansteen}{1997}]{han97} 
Hansteen, V.: 1997, n A. Wilson (ed.), {\it The Corona
and Solar Wind Near Minimum Activity}, Fifth SOHO Workshop, Oslo, 17-20
june 1997, ESA Sp-404, p. 45
\bibitem [\protect\citeauthoryear{Harrison}{1997}]{harr97} 
Harrison, R. A.: 1997, {\it Solar Phys.} {\bf 175}, 457
\bibitem[\protect\citeauthoryear{Harrison {\it et al.}}{1999}]{harr99} 
Harrison, R. A., Lang, J., Brooks, D. H.,
 Innes, D. E: 1999, {\it Astron. Astrphys.} {\bf 351}, 1115
\bibitem[\protect\citeauthoryear{Hasan and Kalkofen}{1999}]{has99} 
Hasan, S.S., and Kalkofen, W.: 1999, {\it Astrophys. J.}
{\bf 519}, 899
\bibitem[\protect\citeauthoryear{Ichimoto and Kurokawa}{1984}]{ich84}
Ichimoto, K. and Kurokawa, K.: 1984,  {\it Solar Phys.}
{\bf 93}, 105
\bibitem[\protect\citeauthoryear{Innes {\it et al.}}{1997}]{inn97}
Innes, D.E., Brekke, P., Germerott, D. and Wilhelm, K.: 1997, {\it Solar Phys.}
{\bf 175}, 341
\bibitem[\protect\citeauthoryear{Krucker {\it et al.}}{1997}]{kru97}
Krucker, S., Benz, A.O., Bastian, T.S. and Acton, L.W.: 1997, {\it
Astrophys. J.} {\bf 488}, 499
\bibitem [\protect\citeauthoryear{Krucker and Benz}{2000}]{kru00} 
Krucker, S. and Benz, A.O.: 2000, {\it Solar Phys.} {\bf 191}, 341
\bibitem [\protect\citeauthoryear{Lites, Rutten, and Kalkofen}{1993}]{lit93}
Lites, B.W., Rutten, R.J. and Kalkofen, W.: 1993, {\it Astrophys. J.} {\bf
414}, 345
\bibitem[\protect\citeauthoryear{Porter {\it et al.}}{1987}]{por87} 
Porter, J.G., Moore, R.L., Reichmann, E.J.,
Engvold, O. and Harvey, K.L.: 1987, {\it Astrophys. J.} {\bf 323}, 380
\bibitem[\protect\citeauthoryear{Scherrer {\it et al.}}{1995}]{scher95} 
Scherrer, P.H.,  et al.: 1995, {\it Solar Phys.} {\bf 162}, 129
\bibitem[\protect\citeauthoryear{Schrijver {\it et al.}}{1997}]{schri97} 
Schrijver, C.J., Title, A.M., van Ballegooijen, A.A., Hagenaar, H.J. 
and Shine, R.A.: 1997, {\it.  Astrophys. J.} {\bf 487}, 424
\bibitem[\protect\citeauthoryear{Shimizu}{1995}]{shim95} Shimizu, T.: 1995, {\it P.A.S.J} {\bf  47}, 251
\bibitem [\protect\citeauthoryear{Thomas, Starr, and Crannell}{1985}]{tho85}
Thomas, R.J., Starr, R., and Crannell, C.J.: 1985, {\it Solar Phys.} {\bf 95}, 323
\bibitem[\protect\citeauthoryear{Ulmschneider, Rosner, and Priest}{1991}]{ulm91}
Ulmschneider, P., Rosner, R., Priest, E.R., eds. 1991, `Mechanisms of
Chromospheric and Coronal Heating', (Berlin:Springer)
\bibitem [\protect\citeauthoryear{Ulrich}{1996}]{ulr96} Ulrich R.K.: 1996, {\it Astrophys. J.} {\bf 465}, 436
\end{thebibliography}
\end{document}